\documentstyle[12pt,epsf,epsfig]{article}
\def\be{\begin{equation}}
\def\ee{\end{equation}}
\def\ba{\begin{eqnarray}}
\def\ea{\end{eqnarray}}

\def\bq{\begin{quote}}
\def\eq{\end{quote}}

 at 10truept

\newcommand{\beq}{\begin{equation}}
\newcommand{\eeq}{\end{equation}}
\newcommand{\beqa}{\begin{eqnarray}}
\newcommand{\eeqa}{\end{eqnarray}}

\def\ltap{\ \raise.3ex\hbox{$<$\kern-.75em\lower1ex\hbox{$\sim$}}\ }
\def\gtap{\ \raise.3ex\hbox{$>$\kern-.75em\lower1ex\hbox{$\sim$}}\ }
\def\gl{\ \raise.5ex\hbox{$>$}\kern-.8em\lower.5ex\hbox{$<$}\ }
\def\roughly#1{\raise.3ex\hbox{$#1$\kern-.75em\lower1ex\hbox{$\sim$}}}

\def\G{{\mathcal G}}
\def\veff{V_{\mbox{\scriptsize{\,eff}}}}
\def\ggb{G^{\mbox{\scriptsize{\,(GB)}}}}

\begin{document}

\begin{titlepage}
\vfill
\begin{flushright}
%\today
\end{flushright}

\vfill
%\vskip 1.0cm
\begin{center}
\baselineskip=16pt
{\Large\bf The Attractor Mechanism in Gauss-Bonnet Gravity}
\vskip 1.0cm
{\large {\sl }}
\vskip 10.mm
{\bf Mohamed M. Anber\footnote{\tt
manber@physics.umass.edu} and
David Kastor\footnote{\tt kastor@physics.umass.edu}} \\
%\\[2mm]
\vskip 1cm
%\vfill
{

       Department of Physics\\
       University of Massachusetts\\
       Amherst, MA 01003\\
}
\vspace{6pt}
\end{center}
\vskip 0.5in
\par
\begin{center}
{\bf ABSTRACT}
\end{center}
\begin{quote}
We study extremal black hole solutions  of $D=5$ Gauss-Bonnet gravity coupled to a system of gauge and scalar fields.  As in Einstein gravity, we find that the values of the scalar fields on the horizon must extremize a certain effective potential  that depends on the black hole charges.  If the matrix of second derivatives of the effective potential at this extremum  has positive eigenvalues, we give evidence, based on a near horizon perturbative expansion, that the attractor mechanism continues to hold in this general class of theories.  We numerically construct solutions to a particular simple single scalar field model that display the attractor mechanism over a wide range of asymptotic values for the scalar field.  We also numerically construct non-extremal solutions and show that the attractor mechanism fails to hold away from extremality.
\vfill
% \hrule width 5.cm
\vskip 2.mm
\end{quote}
\end{titlepage}
%%%%%%%%%%%%%%%%%%%%%%%%%%%%%%%%%%%%%%%%

%
\section{Introduction}

The attractor mechanism
was originally discovered for BPS extremal black holes in  ${\cal N}=2$ supergravity theories \cite{Ferrara:1995ih,Strominger:1996kf,Ferrara:1996dd}.  Under the attractor mechanism, the values of moduli scalars at a BPS extremal black hole horizon  are fixed, independently of their values at infinity,  in terms of the electric and magnetic  charges carried by the black hole. 
More recently, begining with the work of references \cite{Sen:2005wa} and \cite{Goldstein:2005hq}, the attractor mechanism has been investigated for  extremal black holes in non-supersymmetric theories, as well as for non-BPS extremal solutions in ${\cal N}\ge 1$ supersymmetric theories.

The methods employed in references \cite{Sen:2005wa} and \cite{Goldstein:2005hq} are quite different and offer complementary insights into the physics of the attractor mechanism.  Reference  \cite{Sen:2005wa} focuses on the near horizon limit, which is assumed to have geometry $AdS_2\times S^{D-2}$.  An ``entropy function'' is defined by taking the Legendre transform, with respect to the electric charges, of the integral of the Lagrangian density over the $S^{D-2}$.    The constant values of the moduli fields, which solve the equations of motion in the near horizon region, can be shown to be those which extremize this entropy function.  These are the attractor values of the moduli fields.  
Further, the black hole entropy is given by the value of the entropy function at its extremum.
The approach of reference  \cite{Sen:2005wa} is quite general and by design includes the  possibility of higher derivative gravitational interactions.  For instance, equality of the extremum value of the entropy function with the black hole entropy is demonstrated by showing that it reproduces Wald's formula \cite{Wald:1993nt,Jacobson:1993vj,Iyer:1994ys,Jacobson:1994qe} which holds in higher derivative gravity theories. 

In contrast, reference \cite{Goldstein:2005hq} focuses only on Einstein gravity, but obtains very explicit results that illustrate the attractor mechanism in action.   The authors  use a combination of analytic and numerical techniques to follow the radial evolution of the moduli scalars from their attractor values, at an extremal black hole horizon, out to independent values at infinity.  Although this approach lacks the generality of reference \cite{Sen:2005wa}, 
one sees the operation of the attractor mechanism in a vivid way.  
 
In this paper, we will follow the approach of reference \cite{Goldstein:2005hq} and study the attractor mechanism with a  simple higher derivative gravitational interaction, the Gauss-Bonnet term, added to the action.  We focus on $D=5$, the smallest dimension in which the Gauss-Bonnet interaction is non-trivial.  This theory serves as the first example, beyond Einstein gravity, of a Lovelock gravity theory \cite{Lovelock:1971yv}.   Lovelock gravity theories share a number of important  properties with Einstein gravity \cite{Lovelock:1971yv} and have been studied in many contexts over the years.   In particular, vacuum and electrovac black hole solutions have been well studied, begining with the work of \cite{Boulware:1985wk,Wheeler:1985nh,Wheeler:1985qd}.

One knows from the general results of  \cite{Sen:2005wa} that  the moduli scalars  in Gauss-Bonnet gravity must  take values at an extremal black hole horizon that extremize the entropy function for this theory\footnote{The entropy function for Lovelock gravity theories, which includes Gauss-Bonnet gravity, is calculated explicitly in reference \cite{Alishahiha:2006ke} (see also \cite{Prester:2005qs}).}.  Our results establish, at least within the particular gauge and scalar field system we study, that  these near horizon attractor values are actually obtained at the horizon in asymptotically flat, extremal black hole solutions with a range of different values for the scalars at infinity\footnote{Non-supersymmetric, asymptotically flat extremal black hole solutions in a higher derivative gravity theory have also been studied in  reference \cite{Chandrasekhar:2006kx}.  The theory considered in this case is a general $R^2$ gravity theory in $D=4$ with a moduli dependent coupling.  Black hole solutions manifesting the attractor phenomenon are implicitly constructed via a series expansion method.}.

In section (\ref{GBsection}) we describe the $D=5$ Gauss-Bonnet gravity theory coupled to a system of gauge and scalar fields that we will be studying.  In section (\ref{doubleextreme}) we present a set of simple analytic black hole solutions in which the scalar fields take constant values throughout the spacetime.  The possible constant values are, as in reference \cite{Goldstein:2005hq}, the extremal points of a certain effective potential function that depends on the charges carried by the black hole\footnote{See also reference \cite{Ferrara:2006xx} for similar results in $D=5$ Maxwell-Einstein supergravity.}.
In section (\ref{attractormechanism}) we numerically construct extremal solutions in a certain single scalar field model, in which the scalar fields vary between fixed attractor values as the degenerate horizon and independent values at infinity.  These attractor values are again the extremal points of the effective potential, with the provision that the eigenvalues for small fluctuations about the extremal point must all be positive for the attractor mechanism to hold.  We evaluate the ADM mass of these spacetimes and show that it is minimized by the extremal solutions of section (\ref{doubleextreme}) in which the scalars are constant at their attractor values.  Finally, in section (\ref{nonextremal}) we numerically construct non-extremal black hole solutions with non-constant values for the scalar field.  We show that, as in the case of Einstein gravity \cite{Goldstein:2005hq, Astefanesei:2006sy}, non-extremal black holes do not exhibit the attractor mechanism, {\it i.e.} the value of the scalar field at the horizon depends on its value at infinity.

\section{Gauss-Bonnet gravity}\label{GBsection}
We consider Gauss-Bonnet gravity in five dimensions coupled to $N$ Abelian gauge fields $A_\mu^a$ with $a=1,\dots,N$ 
and $n$ massless scalar moduli fields $\phi_{I} $, with $I=1,...,n$.   The moduli scalars have vanishing potential, but couple to the gauge field kinetic terms through the matrix function $f_{ab}(\phi)$.
The action is then given by 
\begin{equation}\label{lagrangian}
S=\frac{1}{\kappa^2}\int\,d x^5\,\sqrt{-g}\,\left[ R+\alpha \,{\cal {L}}^{\mbox{\scriptsize{\,(GB)}}}-2\partial_{\mu} \phi_{I}\partial^{\mu}\phi^{I}-f_{ab}(\phi_J)F^a_{\mu\nu}F^{b\, \mu\nu} \right] \mbox{ ,}
\end{equation} 
with the Gauss-Bonnet term in the Lagrangian given by
\begin{equation}\label{gbterm}
{\cal {L}}^{\mbox{\scriptsize{\,(GB)}}}=R^2-4R_{\gamma\delta}R^{\gamma\delta}+R_{\gamma\delta\lambda\sigma}R^{\gamma\delta\lambda\sigma}\mbox{ ,}
\end{equation}
The coupling constant $\alpha$ has dimensions $(\mbox{length})^2$.  The equations of motion and further details of this theory are given in the Appendix.

We are interested in extremal black hole solutions of this theory.  We will therefore assume a static, spherically symmetric form for the metric
\begin{equation}
\label{spherical metric}
ds^2=-a^{\,2}(r)\,dt^2+\frac{dr^2}{a^{\,2}(r)}+b^{\,2}(r)d\Omega_{3}^{\,2}\mbox{ ,}
\end{equation}
where $d\Omega^2=\gamma_{ij}dx^i dx^j$ with $i,j=1,2,3$ represents the round  metric on the unit $3$-sphere in a general set of coordinates.
We will restrict our attention to black holes that carry only electric charges for the gauge fields $A_\mu^a$.
The equations of motion for the gauge fields may then be solved by taking the  field strengths to be of the form
\begin{equation}\label{gaugefields}
F^a=\frac{f^{ab}\,Q_{b}}{b^3}dt\wedge dr \mbox{.}
\end{equation} 
Here, the constants $Q_a$ are the electric charges and the field dependent tensor $f^{ab}(\phi)$ is the inverse of the tensor coupling $f_{ab}(\phi)$ that appears in the Lagrangian.
With this form for the field strengths, the stress-energy tensor for the gauge fields can be written in terms of  the effective potential 
\begin{equation}
V_{\mbox{\scriptsize{\,eff}}}(\phi)=f^{cd}(\phi)\,Q_{c}\,Q_{d}.
\end{equation}
Furthermore, the effective potential acts as a potential in the equations of motion for the moduli scalars, which are given by 
\begin{equation}\label{scalarequation}
\partial_{r}\left(b^3\,a^2\,\partial_{r}\phi_{I}\right)=
\frac{\partial_{I}V_{\mbox{\scriptsize{\,eff}}}(\phi)}{2b^{\,3}}
\end{equation}
It will be  important to note that these equations of motion may  be solved by constant values $\bar\phi_I$ of the scalar fields, if  these values  represent a  critical point of the effective potential, {\it i.e.} they satisfy
\begin{equation}\label{critical}
(\partial_I\veff ) |_{\bar\phi}=0.
\end{equation}
As in reference  \cite{Goldstein:2005hq}, we will see below that the critical points $\bar\phi_I$ of the effective potential  represent  possible attractor values for the moduli scalars, provided that the matrix of second derivatives of the effective potential
\begin{equation}\label{secondderivs}
M_{IJ}= (\partial_I\partial_J\veff) |_{\bar\phi}
\end{equation}
 has positive eigenvalues.

Given the form of the ansatz for the metric (\ref{spherical metric}) and the gauge fields (\ref{gaugefields}), the gravitational equations of motion are given by
\begin{eqnarray}\label{gravequations}
\nonumber
&&\G_{rr}=(\partial_{r}\phi_{I})(\partial_{r}\phi^{I})-\frac{V_{\mbox{\scriptsize{\,eff}}}(\phi)}{a^2\,b^{\,6}}\mbox{ ,}\qquad \label{aa}
\G_{tt}=a^4\,(\partial_{r}\phi^{I})(\partial_{r}\phi_{I})+\frac{a^2\,V_{\mbox{\scriptsize{\,eff}}}(\phi)}{b^{\,6}}\mbox{ ,}\\
\label{bb}
&&\G_{ij}=\left(-b^{\,2}a^{\,2}(\partial_{r}\phi_{I})(\partial_{r}\phi^{I})+\frac{V_{\mbox{\scriptsize{\,eff}}}(\phi)}{b^{\,4}}\right)\gamma_{ij} \mbox{ ,}
\end{eqnarray}
where the tensor $\G_{\mu\nu}=G_{\mu\nu}+\alpha\,G_{\mu\nu}^{\mbox{\scriptsize{\,(GB)}}}$ combines the Einstein tensor $G_{\mu\nu}$ and its Gauss-Bonnet counterpart $\ggb_{\mu\nu}$ which is given by equation (\ref{counterpart}) in the Appendix, where we also give expressions for the nonzero components of $G_{\mu\nu}$ and $\ggb_{\mu\nu}$ in the spherically symmetric ansatz. 

One particular  combination of the gravitational field equations will be especially important in the analysis below.  Given the gravitational field equations in (\ref{aa}) and the explicit expressions for $G_{\mu\nu}$ and $\ggb_{\mu\nu}$ in the Appendix, 
one can compute the quantity $g^{rr}\G_{rr}-g^{tt}\G_{tt}$ in two ways.  Setting these equal then gives the equation
\begin{equation}\label{linearcomb}
-3{a^2\over b^3}\left[b^2+4\,\alpha\left(1-a\,^2b\,'\,^2\right) \right]b^{\prime\prime} =
2\,a^2\,(\partial_{r}\phi^{I})\partial_{r}\phi_{I}\,,
\end{equation}
which implies that $b^{\prime\prime}=0$ if the scalar fields are constant.  Note that in the absence of the Gauss-Bonnet term, {\it i.e.} with $\alpha=0$, equation (\ref{linearcomb}) implies that $b^{\prime\prime}\le 0$, which is an important ingredient in the `c-theorem' of reference \cite{Goldstein:2005rr}.

\section{Double-extreme solutions}\label{doubleextreme}

Extremal black hole solutions in which the scalar fields take constant values are sometimes called double-extreme solutions.  In this section, we investigate such double-extreme solutions in the Gauss-Bonnet gravity theory of section (\ref{GBsection}).  As noted above, the constant values $\bar\phi_I$ for the scalar fields must be critical points of the effective potential, in order that equation (\ref{scalarequation}) be satisfied.  A second observation is that, with constant values of the scalar fields, equation (\ref{linearcomb}) implies that $b^{\prime\prime}=0$.

\subsection*{Robinson-Bertotti solutions}

We first consider solutions of the Robinson-Bertotti form.  The metric is taken to be 
$AdS_{2}\times S^{3}$, which we write as
\begin{equation}
\label{Robinson-Bertotti metric}
ds^{\,2}=-\frac{x^{\,2}}{R^{\,2}}dt^{\,2}+\frac{R^{\,2}}{x^{\,2}}dx^2+b_{H}^{\,2}\,d\,\Omega_{3}^{2}\,.
\end{equation}
where $R$ and $b_H$ are constants.   This will be the near horizon form of the metric, both for the double-extreme black hole solutions which we study in this section, and the solutions with non-constant scalars which we subsequently study via numerical methods.

First note that since the metric function $b(r)$ in the Robinson-Bertotti metric is constant, we have $b^{\prime\prime}=0$.  Equation (\ref{linearcomb}) then implies that the scalar fields $\phi_I$ must be constant.
The constant values $\bar\phi_I$ of the scalar fields must in turn satisfy equation (\ref{critical}).
Let $\bar V$ be the constant value of the effective potential throughout the spacetime,  $\bar V\equiv \veff(\bar\phi)$.  One can then show that the 
remaining field equations  imply that the $S^3$ and $AdS_2$ radii are given  according to
\begin{equation}\label{RBrelations}
b_H^4={1\over 3}\bar V,\qquad R^2= {1\over 4}b_H^2 +\alpha
\end{equation}
This differs from the Einstein case only in the contribution of the Gauss-Bonnet coupling constant $\alpha$ to the $AdS_2$ radius.

\subsection*{Extremal black hole solutions}
We now consider double extreme black hole solutions\footnote{Because the scalar fields are constant, the black hole solutions we find here are equivalent to those of Gauss-Bonnet gravity coupled to a single $U(1)$ gauge field given in references \cite{Wiltshire:1985us,Wiltshire:1988uq}, with the effective electric charge $Q^{\,2}\propto \bar V$.}.
In order to satisfy $b^{\prime\prime}=0$ with asymptotically flat boundary conditions, we may set $b(r)=r$ without any loss of generality.
Plugging this into the second equation in (\ref{aa}) yields an equation for $a(r)$
\begin{equation}
a\,'\left[3\,a\,r^2+12\,\,\alpha\,a\left(1-a^{\,2} \right) \right]+3r\,a^{\,2}=3r-\frac{\bar V}{r^3}\mbox{ .}
\end{equation}
We can  integrate this equation to obtain the general solution
\begin{equation}\label{generalsolution}
a^2=1+\frac{r^2}{4\,\alpha}\pm\sqrt{\left(1+\frac{r^2}{4\,\alpha}\right)^2-\frac{1}{2 \alpha r^2}\left(r^4-2M r^2+\frac{\bar V}{3}\right)}\,,
\end{equation}
where $M$ is a constant of integration.
One can check that the remaining gravitational equations of motion are also with $a(r)$ having this form.  

This general solution describes both non-extremal and extremal black holes, as well as naked singularities.  Moreover, 
it is well known that Gauss-Bonnet gravity generally has two distinct constant curvature solutions.  If the  cosmological constant vanishes, as it does in the  Lagrangian (\ref{lagrangian}), then flat space is always one of these solutions.  For Gauss-Bonnet coupling $\alpha>0$ ($\alpha<0$) the second constant curvature solution 
has negative (positive) curvature.  We need to sort through these various possibilities to identify the asymptotically flat, extremal black hole solutions in (\ref{generalsolution}).

Taking the large $r$ limit, we see that for $\alpha>0$ the $+$  branch of (\ref{generalsolution}) is asymptotically anti-deSitter, while with $\alpha<0$ the $-$  branch is
asymptotically deSitter, in accordance with the remarks above.  For asymptotically flat solutions, we must then take the $-$ ($+$) branch of (\ref{generalsolution}) for $\alpha>0$ ($\alpha<0$).
Taking the large $r$ limit in these cases leads to
\begin{equation} \label{asquared}
a^2\simeq 1-2\frac{M+\alpha}{r^2}+\frac{\bar V}{3r^4}+\dots \mbox{.}
\end{equation}
Given the normalization of the Einstein term in the Lagrangian (\ref{lagrangian}), the ADM mass can be read off from the expansion 
$g_{tt}\simeq -1 +M_{ADM}/6\pi^2r^2+\dots$.   Comparing with equation (\ref{asquared}) we see that the ADM mass of our solutions is given by $M_{ADM}=12\pi^2(M+\alpha)$.

We must now analyze the horizon structure of the asymptotically flat solutions.  For Gauss-Bonnet coupling $\alpha>0$, taking the $+$ branch in (\ref{generalsolution}), we see that $a^2=0$ when the second term under the square root vanishes.  Horizons then occur at the roots of the polynomial $r^4-2Mr^2 +\bar V/3$.  These roots are given by
\begin{equation}\label{horizons}
r^2_\pm= M\pm\sqrt{M^2-\bar V/3}
\end{equation}
and we see in turn that the solutions represent black holes for $M\ge\sqrt{\bar V/3}$, with the extremality condition being $M=\sqrt{\bar V/3}$.  Keeping track of the signs carefully, one finds that the horizon radii are the same for $\alpha<0$ with one proviso.  In order for the metric function $a^2$ to vanish for $\alpha$ negative, the quantity $1+r^2/4\alpha$ must be negative at the horizon.  For the extremal case of interest to us, this imposes a lower bound on the Gauss-Bonnet coupling constant
\begin{equation}
\alpha\ge -{1\over 4}\sqrt{\bar V/3}.
\end{equation}
Only for $\alpha$ satisfying this bound do we find double extremal black hole solutions. \footnote{Extremal black hole solutions in Gauss-Bonnet gravity are also discussed in references \cite{Neupane:2002bf, Neupane:2003vz}.}

\subsection*{Near horizon limit}

Let us now take the near horizon limit of the double extreme solutions and check that it coincides with the Robinson-Bertotti solutions found at the beginning of the section.  In the extremal limit we find that the outer horizon radius $r_+$ in equation (\ref{horizons})  is given by $r_+^4=M^2=\bar V/3$.  In terms of the parameter $b_H$ of the Robinson-Bertotti solutions, we then have $r_+=b_H$.  The near horizon limit of the metric function $a^2$ in (\ref{generalsolution}) in this case is found to be
\begin{equation}\label{nearhorizon}
a^2\simeq {(r-b_H)^2\over \alpha+ b_H^2/4}.
\end{equation}
Setting $x=r-b_H$ and $R^2=\alpha+ b_H^2/4$, we see that the near horizon limit of the double extreme black hole solutions indeed coincides with the Robinson-Bertotti metric (\ref{Robinson-Bertotti metric}).

\section{Attractor mechanism}\label{attractormechanism}
In the last section, we found extremal black hole solutions with constant scalars in our theory.  As in Einstein gravity \cite{Goldstein:2005hq}, the constant values taken by the scalars $\bar\phi_I$ must represent a critical point of the effective potential $\veff$.  In this section we will construct extremal black hole solutions with non-constant scalars.  These will have the same near horizon limit as the solutions of section (\ref{doubleextreme}).  However, we will see the asymptotic values of the scalar fields may be varied freely.  The existence of these solutions in Gauss-Bonnet gravity establishes the operation  of the attractor mechanism in this theory.

\subsection*{A specific model}

In order to find extremal black hole solutions with non-constant scalars, we must further specify the 
theory by choosing definite numbers of gauge and scalar fields and the form of the couplings between them.
\smallskip
Following reference \cite{Goldstein:2005hq}, we consider a simple example that consists of one scalar field $\phi$ coupled to two $U(1)$ gauge fields $A_\mu^{a}$ with $a=1,2$. The couplings of the scalar field to the gauge fields is taken to be
\begin{equation}
f_{ab}(\phi)=e^{-\alpha_{a}\phi}\delta_{ab}\,.
\end{equation}
It is then straightforward to compute the effective potential for this model, which is given by
\begin{equation}
\label{example of effective potential}
V_{\mbox{\scriptsize{eff}}}(\phi)=e^{\alpha_1\phi}Q_{1}^2+e^{\alpha_2\phi}Q_{2}^2\,.
\end{equation}
In order that a critical point of $\veff$ should exist, the constants $\alpha_1$ and $\alpha_2$ must have opposite signs. The critical value $\bar\phi$  of the scalar field is then given specified by
\begin{equation}
e^{\bar\phi}=\left(\frac{-\alpha_{2}Q_{2} ^2}{\alpha_1Q_{1}^2} \right)^{1/(\alpha_1-\alpha_2)}\,.
\end{equation}
The matrix of second derivatives of the effective potential at the critical point (\ref{secondderivs}) is in the present case simply a number, and is given by
 $M=-2\alpha_1\alpha_2$.  Given that $\alpha_1$ and $\alpha_2$ are assumed to have opposite signs, we see that $M>0$.

\subsection*{Perturbative near horizon analysis}
In their study of the attractor mechanism in Einstein gravity, the authors of \cite{Goldstein:2005hq} were able to follow two routes towards finding solutions with non-constant scalars.  First, they analytically studied solutions that are perturbatively close to double extreme solutions.  In this way, solutions were found in which the asymptotic values of the scalar fields differ by small amounts from their attractor values at the horizon.  Reference \cite{Goldstein:2005hq} also constructs solutions to the exact equations of motion using numerical techniques.  These solutions display the attractor behavior over a wide range of asymptotic values for the scalar field.

In the case of Gauss-Bonnet gravity, the linearized  equations for the scalar field in the background specified by the metric function (\ref{generalsolution}) cannot be solved in closed form\footnote{Note that corrections to the metric functions would not enter until second order in perturbation theory, because of the quadratic nature of the scalar kinetic term and the fact that we are expanding about a critical point of $\veff$.}.  Therefore, to demonstrate the attractor mechanism we will turn very shortly to numerical techniques.   However, in order to appropriately fix  initial conditions for the scalar field $\phi$ near the horizon in our numerical work, we first consider the linearized equation for $\phi$ in this region, which does yield a simple closed form solution.

Although for our numerical work below we will specialize to the single scalar model described above, this near horizon analysis may be carried through in the general case.  Let $\bar\phi_I$ be the critical point values of the scalar fields in a double extreme black hole solution.  
We consider perturbations
\begin{equation}
\phi_{I}(r)=\bar\phi_I+\epsilon\,\phi_{I\,1}(r)\,,
\end{equation}
with $\epsilon\ll 1$.   In order to simplify the analysis, we assume that the fields $\bar\phi_I$ are eigenvectors of the matrix $M_{IJ}$ with corresponding eigenvalues $\beta_I^2$.
In the near horizon regime, we have $b(r)=b_H$ and $a(r)^2=(r-b_H)^2/R^2$.  Setting $x=R-b_H$ as in 
(\ref{Robinson-Bertotti metric}), the linearized scalar field equations are given by
\begin{equation}
x^2\partial^{2}_{x}\phi_{I\,1}+2x\partial_{x}\phi_{I\,1}-\frac{\beta_I^{\,2}\,R^2\,\phi_{I\,1}}{2\,b_H^6}=0\mbox { .}
\end{equation}
The solutions of this equation are given by 
\begin{equation}
\label{solution phi1}
\phi_{I\,1}=C_{I}\left(\frac{x}{b_H}\right)^{\sigma_I^\pm}
\end{equation}
where $\sigma_I^\pm=\left(-1\pm\sqrt{1+2\beta_I^{\,2}\,R^{2}/\,b_H^6}\right)/2$ and the $C_{I}$ are arbitrary constants. 
At this perturbative level, the attractor mechanism works for a given scalar field $\phi_I$, only if the perturbation $\phi_{I\,1}$
vanishes at the horizon $x=0$.  We see that this will be the case if the exponent in (\ref{solution phi1}) is postive \cite{Goldstein:2005hq}.  
The exponent $\sigma_I^-$ is always negative and leads to scalar perturbations that diverge at $x=0$. 
If $\beta_{I}^{\,2}>0$, then the exponent $\gamma_I\equiv \sigma_I^+$ is 
positive and perturbations with this exponent exhibit the attractor mechanism.
However, if $\beta_{I}^{\,2}<0$ then no perturbations exhibiting the attractor behavior exist.
The case $\beta_I^2=0$ is discussed in specific class of examples in reference \cite{Nampuri:2007gv}.  In this case, one must go to higher order in perturbation theory to determine the nature of the scalar perturbations, and whether they exhibit the attractor behavior.
We will assume in the following that all the eigenvalues $\beta_I^2$ of the matrix $M_{IJ}$ are positive.

In order to establish initial conditions for all the relevant degrees of freedom, we also consider the perturbation equations for the metric functions $a^2$ and $b^2$ in the near horizon regime.  As noted above, these begin at second order in perturbation theory, and hence we expand
$a(x)=\bar a(r)+\epsilon^2\,a_2(r)$ and $b(x)=\bar b(r)+\epsilon^2\,b_2(r)$.
From equation (\ref{linearcomb}) we can then obtain the equation for $b_2(x)$
\begin{equation}
\label{ perturbation in b}
\partial_{x}^2 b_2=\frac{-\,b_H^3\sum_{I}\left(\partial_x\phi_{I\,1}\right)^2}{6\,R^{\,2}}\,
\end{equation}
which after insertion of the solution (\ref{solution phi1}) for the first order perturbations to the scalar fields can be integrated to give
\begin{equation}
\label{near horizon b2}
b_2(x)= -\sum_{I}\frac{C_I^2\,b_H^3\,(x/b_H)^{2\,\gamma_I}}{12\gamma_I\,(2\gamma_{I}-1)R^{\,2}}\mbox{ .}
\end{equation}
We see that $b_2(x)$ vanishes at the horizon $x=0$, and hence the metric function $b$ approaches its attractor value $b_H$ at the horizon.

Now one can use the third equation in (\ref{gravequations}) to solve for the purturbation $a_2(x)$ near the horizon and get
\begin{equation}
\partial^2_{x}\left( \bar a\,a_2+\frac{b_H\,\bar a^2}{4\,R^2}\,b_2 \right)=-\frac{b_H^2}{4\,R^2}\left[\frac{\bar a^2\,\partial_x^2 b_2}{b_H}+\frac{4\,b_2\,\bar V}{b_H^7}+\sum_I\left( \bar a^2(\partial_x \phi_{I\,1})^2+\frac{\beta_{I}^2\phi_{I\,1}^2}{2\,b_H^6} \right)\right]\,.
\end{equation}
The integration of this equation gives
\begin{eqnarray}
\nonumber
\bar a\,a_2+\frac{b_H\,\bar a^2}{4\,R^2}\,b_2&=&-\sum_{I}\frac{b_H^2\,C_I^2}{4\,R^2(2\gamma_I+1)(2\gamma_I+2)b_H^{2\gamma_I}}\times\\
\label{near horizon a2}
&\times&\left[ (1-\frac{b_H^2}{6\,R^2})\frac{\gamma_I^2}{R^2}-\frac{\bar V}{3\,b_H^4\,\gamma_{I}(2\gamma_I-1)R^2}+\frac{\beta_I^2}{2\,b_H^6}\right]x^{2\gamma_I+2}\,,
\end{eqnarray}
from which we see that the perturbation to $a(r)$ also vanishes as $x\rightarrow 0$, consistent with the attractor phenomenon.
\subsection*{Numerical solutions}
We now return to the single scalar field model and carry out a numerical analysis of solutions to the 
full nonlinear field equations.  From the $ij$ components of the field equations in (\ref{gravequations}) together  with (\ref{linearcomb}), we obtain the radial evolution equations
\begin{eqnarray}
\nonumber
\phi\,''(r)&=&-(3\,\frac{b\,'}{b}+2\,\frac{a\,'}{a})\phi\,'+\frac{\partial_{\phi}V_{\mbox{\scriptsize{eff}}}(\phi)}{2\,a^{\,2}\,b^{\,6}}\\
\nonumber
b\,''(r)&=&\frac{-2\,b^{\,3}}{3}\frac{\phi\,'^2}{b^{\,2}+4\,\alpha(1-a^{2}\,b\,'^{\,2})}\\
\nonumber
a\,''(r)&=&\frac{1}{a\left[b^{\,2}+4\alpha(1-a^{\,2}b\,'^{\,2})\right]}\left[1-a^{\,2}\,b\,'^{\,2}-b^{\,2}\,a\,'^{\,2}-2\,a\,b\,\left(2\,a\,'\,b\,'+a\,b\,''\right)\right.\\
\label{explicit nonlinear}
&-&\left.4\,\alpha\,a\,'^{\,2}\left(1-3\,a^{\,2}b\,'^{\,2}\right)+8\,\alpha\,a^{\,3}\,a\,'\,b\,'\,b\,''-b^{\,2}\,a^{\,2}\left(\partial_{r}\phi\right)^{\,2}+\frac{V_{\mbox{\scriptsize{eff}}}(\phi)}{b^{\,4}} \right]\,.
\end{eqnarray}
We integrate these equations numerically  using the Rung-Kutta method. As discussed in \cite{Goldstein:2005hq}, one cannot impose boundary conditions near $r=\infty$, because  the growing mode of the scalar field in (\ref{solution phi1}) will lead to divergent results near the horizon. Instead, we integrate outward in radius.  We start the numerical integration at an initial point $r_i$ close to the horizon and use the perturbative near horizon results in (\ref{solution phi1}), (\ref{near horizon b2}) and (\ref{near horizon a2}) to fix  initial conditions there. We denote the proximity to the horizon by the parameter
\begin{equation}
\delta r=\frac{r_i-b_H}{r_i}\,.
\end{equation}
With a single scalar field, the strength of the perturbation near the horizon is determined by the choice of the single constant $C$ in equation (\ref{solution phi1}).

As we integrate the system of equations,  we check that the numerical solution satisfies the constraint given by the $rr$ component of the field equations (\ref{gravequations}) which contains no second order
radial derivatives
\begin{equation}
a\,'(r)=a\,b^{\,3}\,\left[\frac{3}{a^{\,2}\,b^{\,2}}\left(1-a^{\,2}\,b\,'^{\,2}\right)+\left(\phi\,'\right)^2-\frac{V_{\mbox{\scriptsize{eff}}}(\phi)}{a^{\,2}\,b^{\,6}}\right]/3\,b\,'\,\left[b^{\,2}+4\,\alpha\,(1-a^{2}\,b\,'^{\,2})\right]\,.
\end{equation}
We find that in all cases the constraint is indeed satisfied to appropriate numerical accuracy.

As a further check, we have also numerically integrated the simpler lowest order perturbative equations for  $\phi_1(r)$, $a_2(r)$ and $b_{\,2}(r)$ expanded about the full, asymptotically flat,  double extreme black hole background.  These perturbative results should be accurate for sufficiently small values of the parameter $C$.  We can then check that our results for integrating the full nonlinear field equation match up with the perturbative results for small $C$.

\begin{figure}[ht]
\leftline{
\includegraphics[width=0.5\textwidth]{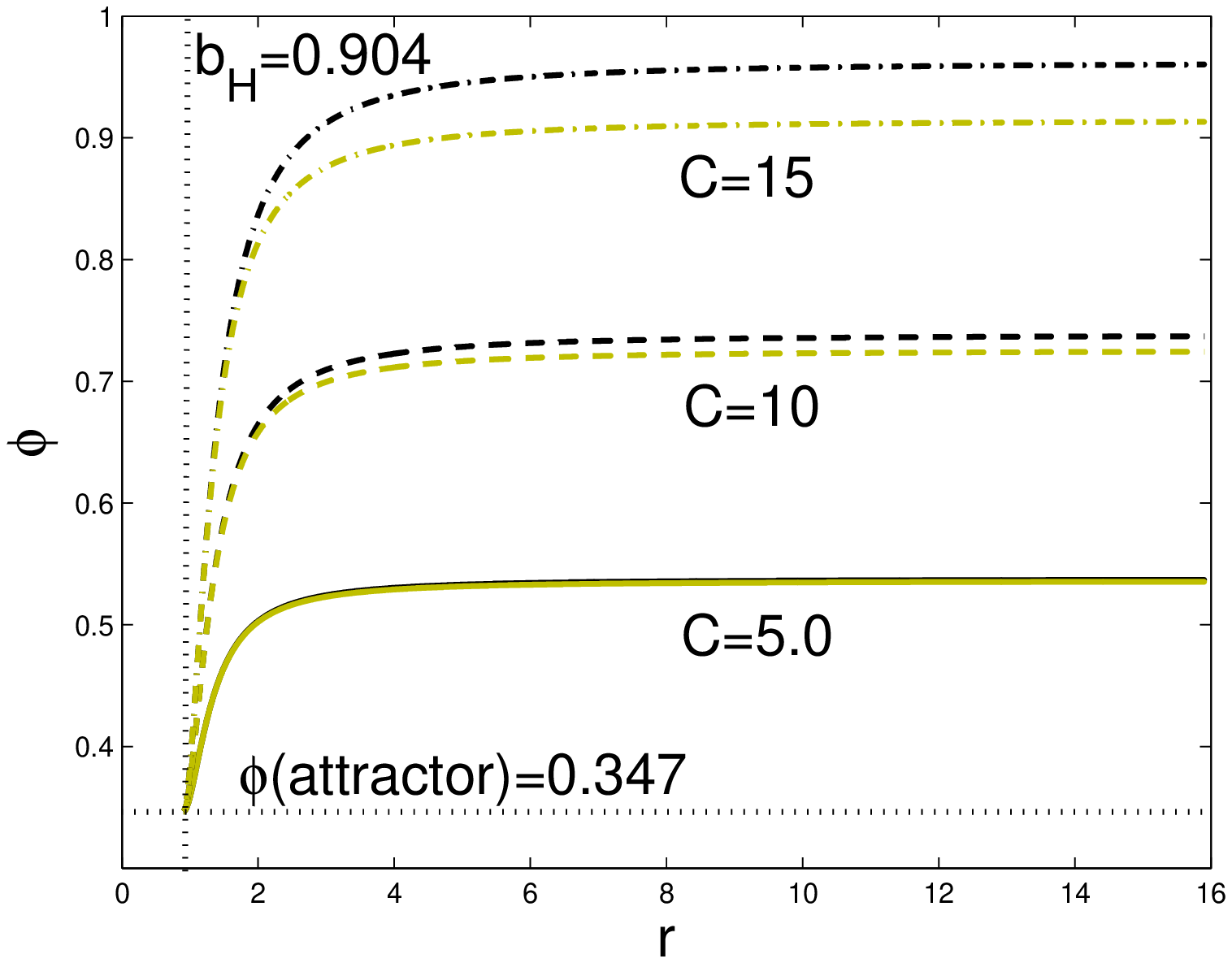}
\includegraphics[width=0.5\textwidth]{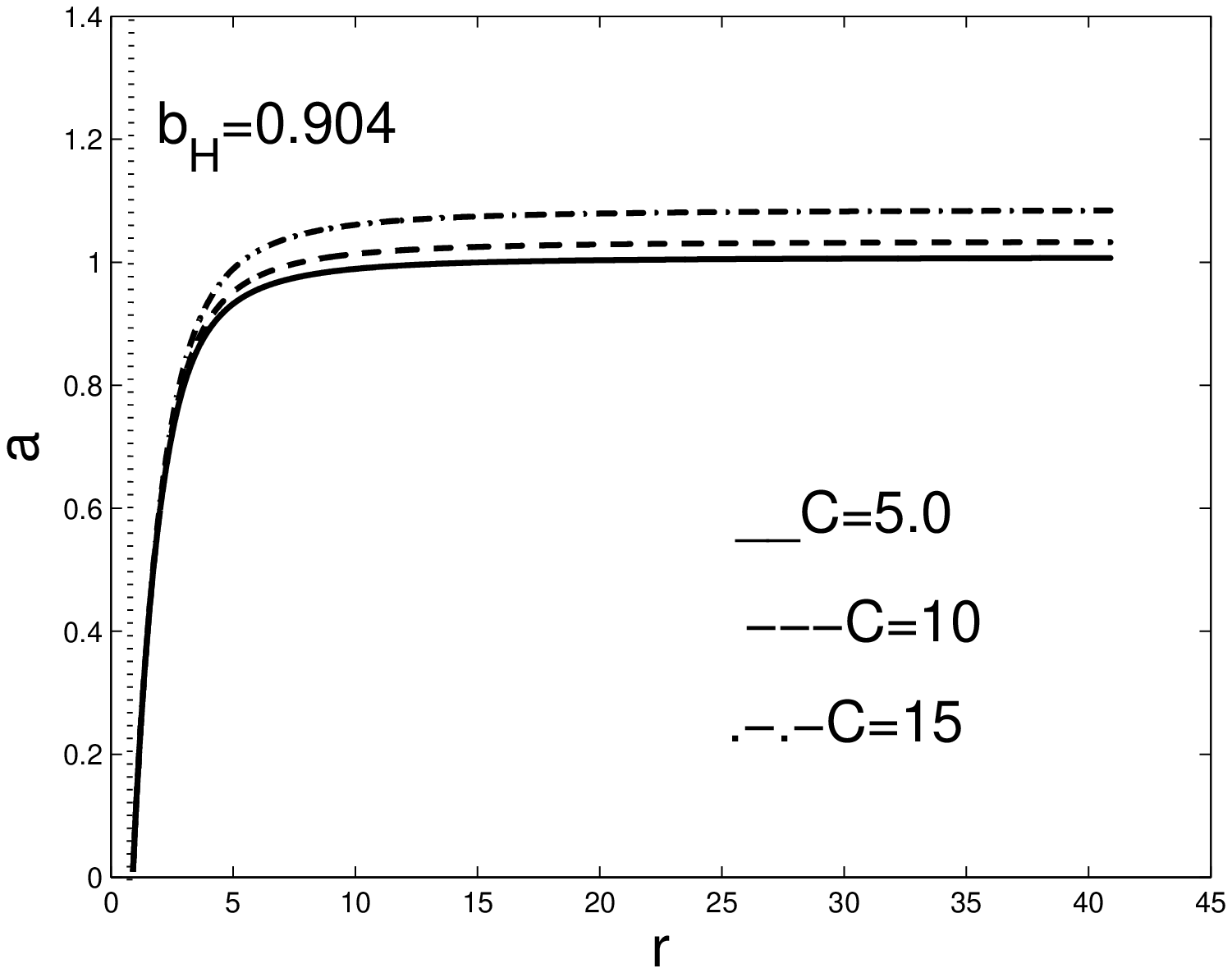}
}
\leftline{
\includegraphics[width=0.5\textwidth]{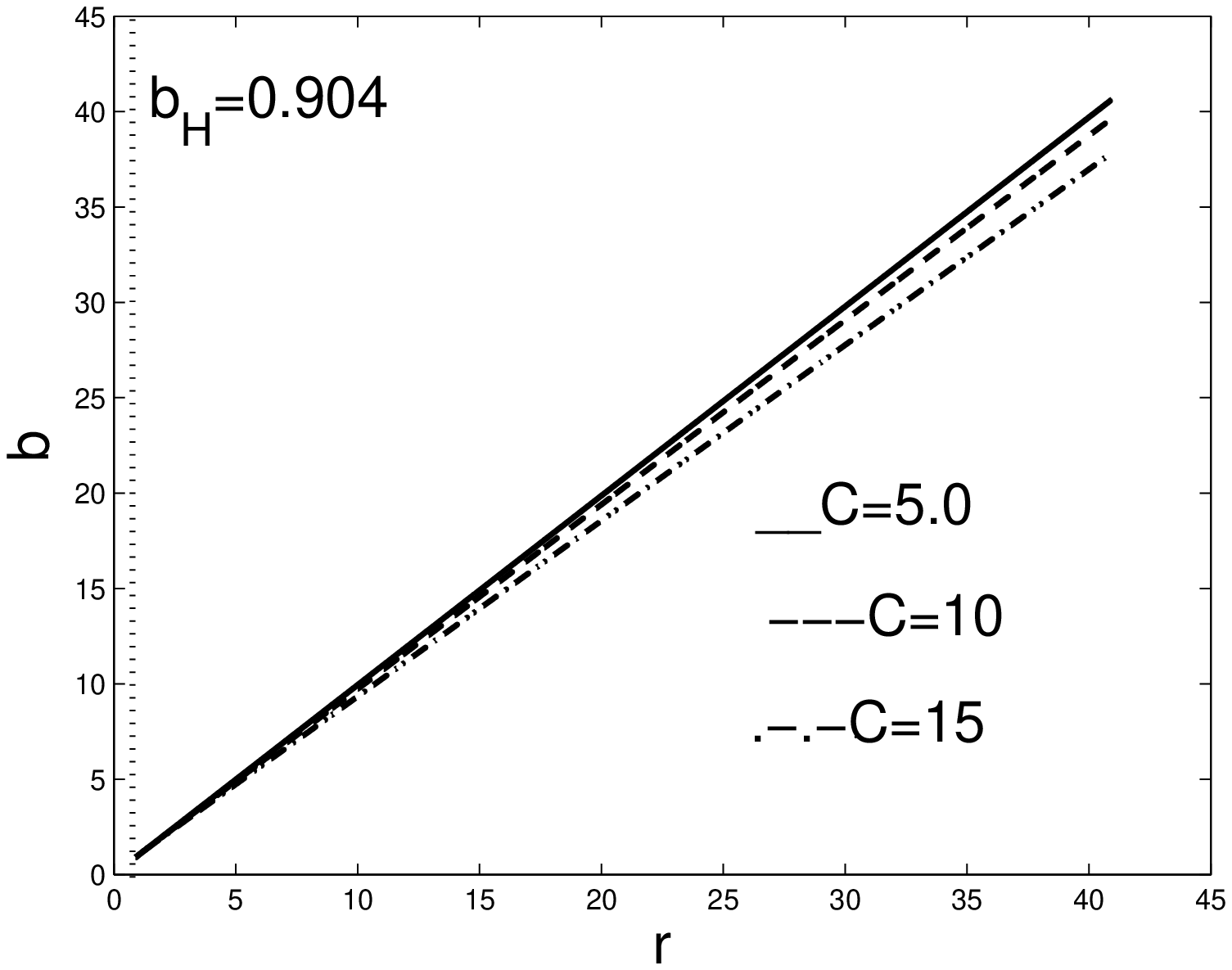}
}
\caption{This figure displays the results of numerically integrating the field equations with 
$C =\{5.0,10,15\}$.  The parameters of the scalar field model are taken to be $\alpha=1.0$, 
$Q_{1}=1/\sqrt 2$ and $Q_{2}=\sqrt2$ , $\alpha_1=-\alpha_2=2.0$, which gives $b_H=0.904$ and $\bar\phi=0.347$.  The numerical integrations are started at $\delta r=0.01$.
In the plot of $\phi$, the dark lines are the result of integrating the full nonlinear field equations, while the lighter lines show the first order perturbative results.  We see that these results agree well at the smallest value of $C$ displayed.  The graphs for $a$ and $b$ show that the exact solutions are singularity free and asymptotically flat.
}
\label{exact figure}
\end{figure}

Our numerical results are displayed in figure (\ref{exact figure}).  The vertical line in the plots denotes the horizon radius $b_H$.   The plot of $\phi(r)$ clearly shows the attractor behavior.  By construction, the plots start near the horizon with small deviations determined by the value of $C$ from the attractor value $\bar\phi$, which is indicated by the horizontal line on the plot.  We see that for large radius, the scalar field $\phi$ approaches constant values that can differ quite significantly from its attractor value near the horizon.  The $\phi$ plot also includes for comparison the results of integrating the first order perturbative equation for $\phi$.  We see that these results have the same qualitative features as the results from the full noninear equations and that the two sets of results have good quantitative agreement at the smallest value of $C$ displayed.

The plot of the metric functions $a(r)$ and $b(r)$ have forms consistent with asymptotic flatness, with $a(r)$ going to a constant and $b(r)$ growing linearly at large radius.  We will see below that their detailed asymptotic behavior gives asymptotically flat solutions with finite ADM mass.

\subsection*{Black hole mass}
In this section we evaluate the ADM mass for our numerical solutions.  
It is a basic feature of the supersymmetric attractor mechanism that the ADM mass is minimized, for fixed values of the electromagnetic charges carried by the black hole, when the scalar fields take their attractor values throughout the spacetime \cite{Ferrara:1996dd}.  
It was shown in \cite{Goldstein:2005hq} that this continues to be the case for nonsupersymmetric attractors in $4$-dimensional Einstein gravity.  In the following we show that this behaviour 
persists, at least in our $5$-dimensional example, when the Gauss-Bonnet interaction term is added to the gravitational Lagrangian. 

As noted above, our numerical results show that at large radius $b(r)$ increases linearly with radius and $a(r)$ approaches a constant value.  If we assume that the approach of $a(r)$ to its asymptotic value near infinity is power law with some exponent, then for large $r$ we have
\begin{equation}\label{at infinity}
b(r)=f\,r\,,\qquad  a^{\,2}(r)=a_{\infty}^{\,2}-\frac{M_{ADM}}{6\,\pi^{\,2}\,r^{n}}\,,
\end{equation}
where $f$, $a_{\infty}$ and $n$ are constants.  For an asymptotically flat spacetime in $5$-dimensions, we should have $n=2$, and $M_{ADM}$ in (\ref{at infinity}) would then be 
the properly normalized ADM mass.
Rescaling the time coordinate so that $g_{tt}$ approaches $-1$ at infinity, the asymptotic form of the metric is then
\begin{equation}
ds^{2}=-\left(1-\frac{ M_{ADM}\,f^{\,n}}{6\,\pi^{\,2}\,a_{\infty}^{\,2}y^{\,n}} \right)dt^{2}+\frac{d\,y^{\,2}}{a_{\infty}^{2}\,f^{2}\left(1-\frac{M_{ADM}\,f^{\,n}}{6\,\pi^{\,2}\,a_{\infty}^{\,2}y^{\,n}} \right)}+y^{\,2}d\Omega_{3}^{\,2}\,.
\end{equation}
where $y=fr$.
It is straightforward to check that the constants $f$ and $a_\infty$ satisfy the relation $f a_\infty=1$ in our numerical solutions.  In order to check that $n=2$ in our solutions and to determine the value of 
$M_{ADM}$, we made a log vs. log plot of $-(a^2-a_\infty^2)$ versus $y$.
Via the relation 
\begin{equation}
\label{loglog equation}
\log\left(a_{\infty}^{\,2}-a^{\,2} \right)=\log\left(\frac{M_{ADM}\,f^{\,n}}{6\,\pi^{\,2}} \right)-n\,\log\left(y\right)\,,
\end{equation}
we see that for power law falloff of $a(r)$ to its asymptotic value this plot should approach a straightline with slope $-n$.  In this manner, we determined that $n=2$ in our numerical solutions to good accuracy and obtained values of $M_{ADM}$ over a range of values of $C$.  

\begin{figure}[ht]\label{massfig}
\leftline{
\includegraphics[width=0.5\textwidth]{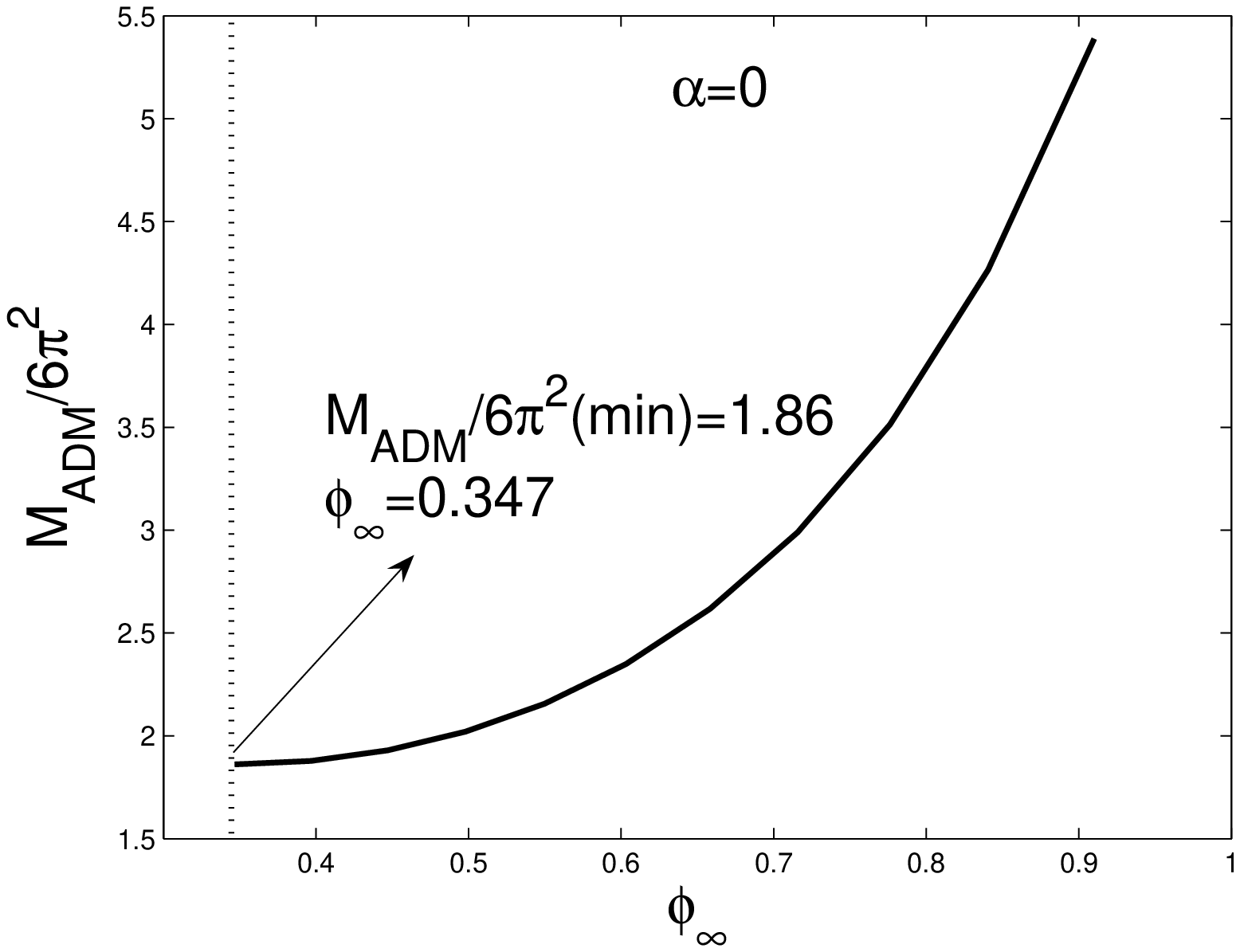}
\includegraphics[width=0.5\textwidth]{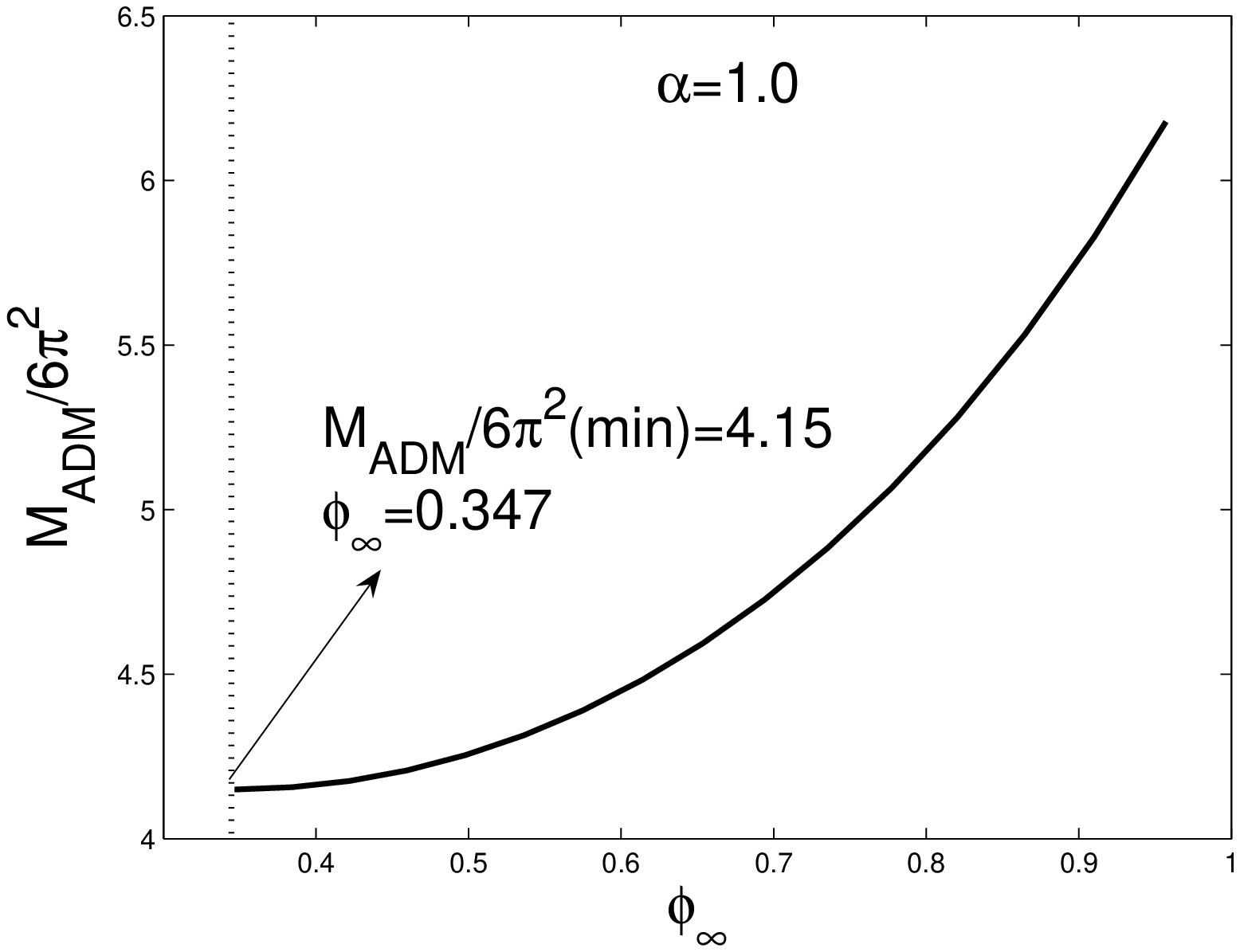}
}
\caption{The black hole mass $M_{ADM}$ versus the asymptotic value of the scalar field $\phi_{\infty}$ for $\alpha=0$ and $\alpha=1.0$. We chose charges $Q_{1}=1/\sqrt 2$ and $Q_{2}=\sqrt2$ , $\alpha_1=-\alpha_2=2.0$ and $\delta r=0.01$. Plots show that mass increases with $\phi_{\infty}$. The minimum value of mass is that of the double-extreme black hole obtained by setting $\phi_{\infty}$ equal to its critical value at the horizon.}
\label{black hole Mass}
\end{figure}

Figure (\ref{massfig}) shows plots of $M_{ADM}$ with varying asymptotic values for $\phi$ for 
$\alpha=0$ ({\it i.e.} pure Einstein gravity) and $\alpha=1$.   We see that the results are qualitatively similar for the two values of $\alpha$ and that in both cases the mass is minimized when the scalar field takes its attractor value $\bar\phi$ throughout the spacetime.

\section{Non-extremal black holes}\label{nonextremal}
Finally, we consider nonextremal black hole solutions.  We saw in section (\ref{doubleextreme}) that 
nonextremal solutions exist with the scaler fields fixed at a critical point of 
 $V_{\mbox{\scriptsize{eff}}}(\phi)$. The metric functions are given by 
\begin{eqnarray}
a^{\,2}(r)=1+\frac{r^{\,2}}{4\,\alpha}-\sqrt{\left(1+\frac{r^{\,2}}{4\,\alpha}\right)^{2}-\frac{\left(r^{\,2}-r_{+}^{\,2} \right)\left(r^{\,2}-r_{-}^{\,2} \right)}{2\,\alpha\,r^{2}}}\,, && b(r)=r\,,
\end{eqnarray}
where $r_{\pm}$ are the inner and outer horizon radii with
\begin{eqnarray}
r^{2}_{\pm}=M\pm\sqrt{M^{\,2}-\frac{V_{\mbox{\scriptsize{eff}}}(\phi_{i0})}{3}}\,.
\end{eqnarray}
We want to ask whether there exist nonextremal attractor solutions in which the scalar fields vary between their attractor values at the outer horizon and independent values at infinity?  In the Einstein case \cite{Goldstein:2005hq, Astefanesei:2006sy} such solutions do not exist, and we expect to find similar results after adding the Gauss-Bonnet interaction.  We address this question both analytically, by lookiing at perturbations to the scalar field in the near horizon region, and numerically by looking at solutions to the full nonlinear field equations.

\begin{figure}[ht]
\leftline{
\includegraphics[width=0.5\textwidth]{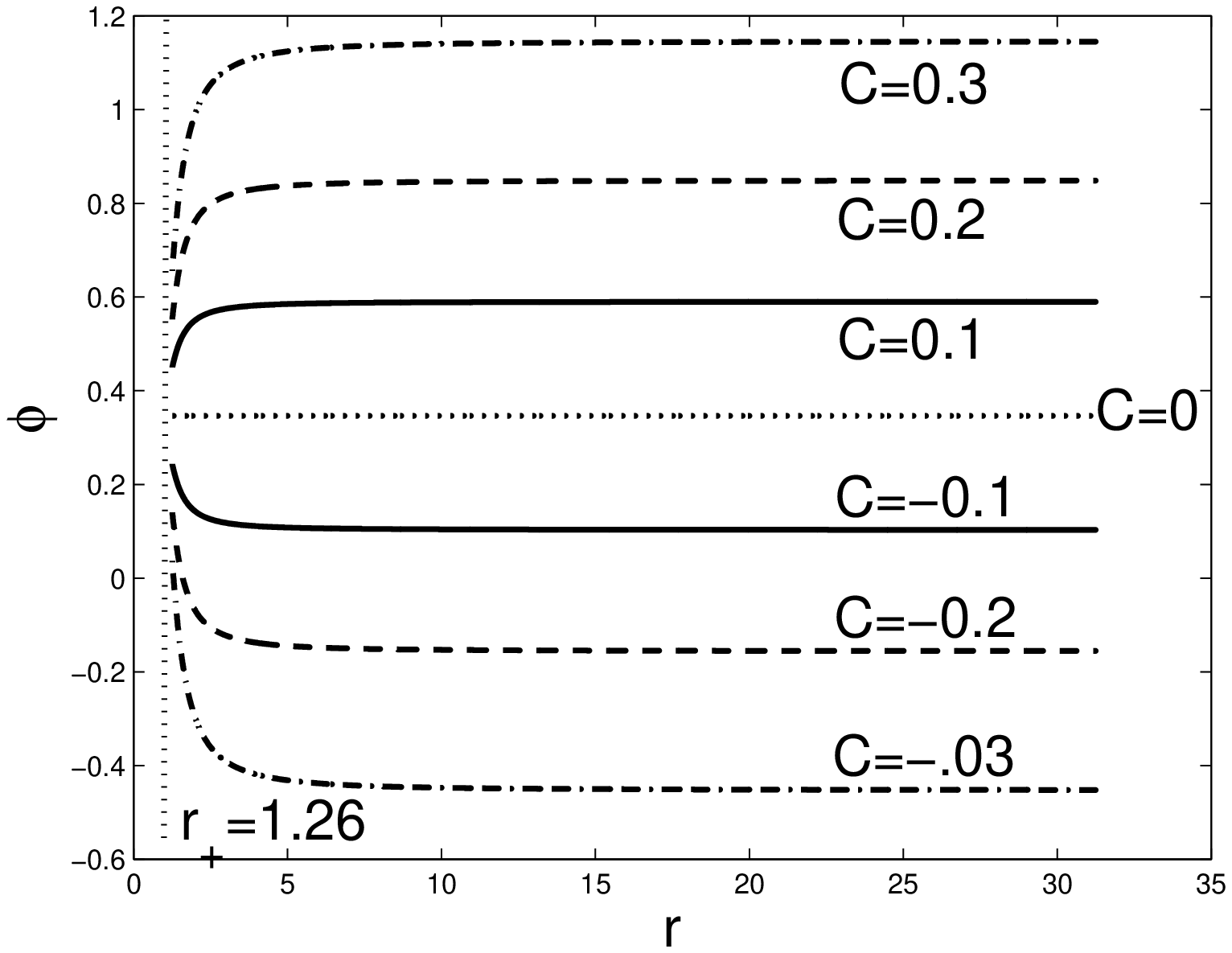}
}
\caption{The non-extremal black hole . We choose $\alpha=1.0$ $M=1.0$, charges $Q_{e1}=1/\sqrt 2$ and $Q_{e2}=\sqrt2$ , $\alpha_1=-\alpha_2=2.0$ and $\delta r=0.01$. Plot shows that the non-extremal solution does not exhibit the attractor mechanism. The scalar field is drawn to different values at the horizon for different values at infinity.  }
\label{nonextermal blach hole}
\end{figure}

We give the perturbative results first.
Near the outer horizon the leading order behavior of the metric functions expanded in terms of  $r-r_{+}$ is given by 
\begin{equation}
a^{\,2}(r)\simeq\rho(r_{+},r_{-})(r-r_{+})\qquad b(r)\simeq r_+ 
\end{equation}
where $\rho(r_{+},r_{-})=2(r_{+}^{\,2}-r_{-}^{\,2})/(4\,\alpha\,r_{+}+r_{+}^{\,3})$. 
The first order perturbative equation for the scalar field $\phi$ in the near horizon region is then given by
\begin{equation}
(r-r_{+})\phi_{i1}\,''+\phi_{i1}\,'-\frac{\beta^{2}}{2\,r_{+}^{\,6}\,\rho}\phi_{i\,1}=0\,,
\end{equation}
where $\beta^{\,2}=\partial_{i}\partial_{j}V_{\mbox{\scriptsize{eff}}}(\phi_{i0})$. The solutions for linearized perturbations of the scalar field are then given by
\begin{equation}
\label{nonextermalphi}
\phi_{1}(r)=C\,I_{0}\left[\frac{\beta}{r_{+}^{\,3}}\sqrt{\frac{2(r-r_{+})}{\rho}} \right]+D\,K_{0}\left[\frac{\beta}{r_{+}^{\,3}}\sqrt{\frac{2(r-r_{+})}{\rho}} \right]\,,
\end{equation}
where $I_{0}$ and $K_{0}$ are the modified Bessel's functions of the first and second kind respectively. Since $K_0$ is singular at $r=r_+$ we set $D=0$. However, we also have $I_0(x=0)\neq 0$.   small perturbations to the scalar field therefore necessarily modify its value at the horizon, and we see that the attractor mechanism no longer holds for nonextremal black hole horizons.
Figure (\ref{nonextermal blach hole}) displays our numerical results for nonextremal solutions to the full nonlinear field equations for different values of the parameter $C$.  The fact that the values for $\phi$ at the horizon differ from the attractor values agrees with the result of the perturbative analysis and shows that the near horizon results extend to nonextremal asymptotically flat solutions.

\subsection*{Acknowledgements}

This work was supported in part by NSF grant PHY-0555304.

\section*{Appendix}

In this appendix, we present further details of the formalism of section (\ref{GBsection}).  The Lagrangian is given in equations (\ref{lagrangian}) and (\ref{gbterm}).  The equations of motion for the metric, gauge and moduli fields are given by
\begin{equation}
\G_{\mu\nu}=2(\partial_{\mu}\phi_{I})\partial_{\nu}\phi^{I}-g_{\mu\nu}(\partial_{\alpha}\phi_{I})\partial^{\alpha}\phi^{I}+f_{ab}(\phi)\left[2F^{a}_{\mu\lambda}F^{b\,\,\,\lambda}_{\nu}-\frac{1}{2}g_{\mu\nu}F^{a}_{\mu\nu}F^{b\,\mu\nu} \right]\mbox{ ,}
\end{equation}
\begin{equation}
\frac{1}{\sqrt{-g}}\partial_{u}\left(\sqrt{-g}\partial^{\mu}\phi_{I} \right)=\frac{1}{4}\partial_{I}(f_{ab})F^{a}_{\mu\nu}F^{b\,\mu\nu} \mbox{ ,}
\end{equation}
\begin{equation}
\partial_{\mu}\left(\sqrt{-g}f_{ab}F^{b\,\mu\nu} \right)=0\mbox{ ,}
\end{equation}
with $\G_{\mu\nu}=G_{\mu\nu}+\alpha\,G_{\mu\nu}^{\mbox{\scriptsize{\,(GB)}}}$, where $G_{\mu\nu}$ is the Einstein tensor and $G_{\mu\nu}^{\mbox{\scriptsize{\,(GB)}}}$ is its Gauss-Bonnet counterpart given by
\begin{equation}\label{counterpart}
G_{\mu\nu}^{\mbox{\scriptsize{\,(GB)}}}=2\left(R_{\mu\sigma\kappa\tau}R_{\nu}^{\,\,\,\sigma\kappa\tau}-2R_{\mu\rho\nu\sigma}R^{\rho\sigma}-2R_{\mu\sigma}R^{\sigma}_{\,\,\,\nu}+RR_{\mu\nu} \right)-\frac{1}{2}g_{\mu\nu}{\cal {L}}^{\mbox{\scriptsize{\,(GB)}}} \mbox{ .}
\end{equation}
The nonzero components of the Riemann tensor for the spherically symmetric metric  (\ref{spherical metric}) are given by 
\begin{eqnarray}\label{one}
R_{rtr}{}^{t}&=&-({a^{\prime\prime}\over a}+{a^{\prime\; 2}\over a^2}),\quad
R_{rir}{}^{j}= - ({b^{\prime\prime}\over b} + {a^\prime b^\prime\over ab})\delta_i^j,\\
R_{tit}{}^j&=& a^4({a^\prime b^\prime\over ab})\delta_i^j,\quad
R_{ijk}{}^l=(1-a^2b^{\prime\; 2})(\gamma_{ik}\gamma_j{}^l-\gamma_{jk}\gamma_i{}^l)\label{two}
\end{eqnarray}
The nonzero components of the Einstein tensor $G_{\mu\nu}$ and its Gauss-Bonnet counterpart $\ggb_{\mu\nu}$ are given by
\begin{eqnarray}\label{three}
%\nonumber
G_{tt}&=&{3a^2\over b^2}(1-a^2b^{\prime\; 2})-3a^4({b^{\prime\prime}\over b}+ 
{b^{\prime\; 2}\over b^2})\mbox{, }\qquad 
G_{rr}= -3{1\over a^2b^2}(1-a^2b^{\prime\; 2}) + 3 {a^\prime b^\prime\over ab}
\mbox{, }\\ \label{four}
%\nonumber
G_{ij}&=&\left(  -(1-a^2b^{\prime\; 2}) +b^2\, a^2({a^{\prime\prime}\over a}+ 
{a^{\prime\; 2}\over a^2}) +2a^2b^2({b^{\prime\prime}\over b}+ 2
{a^\prime b^\prime\over ab})
\right)\gamma_{ij}\mbox { ,}
\end{eqnarray}
while those of its Gauss-Bonnet counterpart are given by
\begin{eqnarray}
%\nonumber
%G_{tt}^{\mbox{\scriptsize{\,(GB)}}}
\ggb_{tt}&=& -12{a^4\over b^2}(1-a^2b^{\prime\; 2})
({b^{\prime\prime}\over b}+ {a^\prime b^\prime\over ab})
\mbox { ,}\qquad
%G_{rr}^{\mbox{\scriptsize{\,(GB)}}}
\ggb_{rr}= 12 {1\over b^2}(1-a^2b^{\prime\; 2})({a^\prime b^\prime\over ab})
\mbox{ ,}\\
\label{all components}
%G_{ij}^{\mbox{\scriptsize{\,(GB)}}}
\ggb_{ij}&=& 4a^2\,\left\{(1-a^2b^{\prime\; 2})({a^{\prime\prime}\over a}+ 
{a{\prime\; 2}\over a^2}) - 2a^2b^2({a^\prime b^\prime\over ab}) ({b^{\prime\prime}\over b}+ {a^\prime b^\prime\over ab})\right\}\gamma_{ij}
 \mbox{ .}
\end{eqnarray}


\begin{thebibliography}{99}

%\cite{Ferrara:1995ih}
\bibitem{Ferrara:1995ih}
  S.~Ferrara, R.~Kallosh and A.~Strominger,
  ``N=2 extremal black holes,''
  Phys.\ Rev.\  D {\bf 52}, 5412 (1995)
  [arXiv:hep-th/9508072].
  %%CITATION = PHRVA,D52,5412;%%
  
%\cite{Strominger:1996kf}
\bibitem{Strominger:1996kf}
  A.~Strominger,
  ``Macroscopic Entropy of $N=2$ Extremal Black Holes,''
  Phys.\ Lett.\  B {\bf 383}, 39 (1996)
  [arXiv:hep-th/9602111].
  %%CITATION = PHLTA,B383,39;%%
  
%\cite{Ferrara:1996dd}
\bibitem{Ferrara:1996dd}
  S.~Ferrara and R.~Kallosh,
  ``Supersymmetry and Attractors,''
  Phys.\ Rev.\  D {\bf 54}, 1514 (1996)
  [arXiv:hep-th/9602136].
  %%CITATION = PHRVA,D54,1514;%%
  
  %\cite{Sen:2005wa}
\bibitem{Sen:2005wa}
  A.~Sen,
  ``Black hole entropy function and the attractor mechanism in higher
  derivative gravity,''
  JHEP {\bf 0509}, 038 (2005)
  [arXiv:hep-th/0506177].
  %%CITATION = JHEPA,0509,038;%%


%\cite{Goldstein:2005hq}
\bibitem{Goldstein:2005hq}
  K.~Goldstein, N.~Iizuka, R.~P.~Jena and S.~P.~Trivedi,
  ``Non-supersymmetric attractors,''
  Phys.\ Rev.\  D {\bf 72}, 124021 (2005)
  [arXiv:hep-th/0507096].
  %%CITATION = PHRVA,D72,124021;%%


%\cite{Wald:1993nt}
\bibitem{Wald:1993nt}
  R.~M.~Wald,
  ``Black hole entropy is the Noether charge,''
  Phys.\ Rev.\  D {\bf 48}, 3427 (1993)
  [arXiv:gr-qc/9307038].
  %%CITATION = PHRVA,D48,3427;%%
  
  %\cite{Jacobson:1993vj}
\bibitem{Jacobson:1993vj}
  T.~Jacobson, G.~Kang and R.~C.~Myers,
  ``On Black Hole Entropy,''
  Phys.\ Rev.\  D {\bf 49}, 6587 (1994)
  [arXiv:gr-qc/9312023].
  %%CITATION = PHRVA,D49,6587;%%
  
  %\cite{Iyer:1994ys}
\bibitem{Iyer:1994ys}
  V.~Iyer and R.~M.~Wald,
  ``Some properties of Noether charge and a proposal for dynamical black hole
  entropy,''
  Phys.\ Rev.\  D {\bf 50}, 846 (1994)
  [arXiv:gr-qc/9403028].
  %%CITATION = PHRVA,D50,846;%%
  
  %\cite{Jacobson:1994qe}
\bibitem{Jacobson:1994qe}
  T.~Jacobson, G.~Kang and R.~C.~Myers,
  ``Black hole entropy in higher curvature gravity,''
  arXiv:gr-qc/9502009.
  %%CITATION = GR-QC/9502009;%%
  
  %\cite{Lovelock:1971yv}
\bibitem{Lovelock:1971yv}
  D.~Lovelock,
  ``The Einstein tensor and its generalizations,''
  J.\ Math.\ Phys.\  {\bf 12}, 498 (1971).
  %%CITATION = JMAPA,12,498;%%
  
  %\cite{Boulware:1985wk}
\bibitem{Boulware:1985wk}
  D.~G.~Boulware and S.~Deser,
  ``String Generated Gravity Models,''
  Phys.\ Rev.\ Lett.\  {\bf 55}, 2656 (1985).
  %%CITATION = PRLTA,55,2656;%%
  
  %\cite{Wheeler:1985nh}
\bibitem{Wheeler:1985nh}
  J.~T.~Wheeler,
  ``Symmetric Solutions To The Gauss-Bonnet Extended Einstein Equations,''
  Nucl.\ Phys.\  B {\bf 268}, 737 (1986).
  %%CITATION = NUPHA,B268,737;%%
  
  %\cite{Wheeler:1985qd}
\bibitem{Wheeler:1985qd}
  J.~T.~Wheeler,
  ``Symmetric Solutions To The Maximally Gauss-Bonnet Extended Einstein
  Equations,''
  Nucl.\ Phys.\  B {\bf 273}, 732 (1986).
  %%CITATION = NUPHA,B273,732;%%
  
  %\cite{Alishahiha:2006ke}
\bibitem{Alishahiha:2006ke}
  M.~Alishahiha and H.~Ebrahim,
  ``Non-supersymmetric attractors and entropy function,''
  JHEP {\bf 0603}, 003 (2006)
  [arXiv:hep-th/0601016].
  %%CITATION = JHEPA,0603,003;%%

%\cite{Prester:2005qs}
\bibitem{Prester:2005qs}
  P.~Prester,
  ``Lovelock type gravity and small black holes in heterotic string theory,''
  JHEP {\bf 0602}, 039 (2006)
  [arXiv:hep-th/0511306].
  %%CITATION = JHEPA,0602,039;%%
  
  
   %\cite{Chandrasekhar:2006kx}
\bibitem{Chandrasekhar:2006kx}
  B.~Chandrasekhar, S.~Parvizi, A.~Tavanfar and H.~Yavartanoo,
  ``Non-supersymmetric attractors in R**2 gravities,''
  JHEP {\bf 0608}, 004 (2006)
  [arXiv:hep-th/0602022].
  %%CITATION = JHEPA,0608,004;%%
  
  %\cite{Ferrara:2006xx}
\bibitem{Ferrara:2006xx}
  S.~Ferrara and M.~Gunaydin,
  ``Orbits and attractors for N = 2 Maxwell-Einstein supergravity theories in
  five dimensions,''
  Nucl.\ Phys.\  B {\bf 759}, 1 (2006)
  [arXiv:hep-th/0606108].
  %%CITATION = NUPHA,B759,1;%%

%\cite{Astefanesei:2006sy}
\bibitem{Astefanesei:2006sy}
  D.~Astefanesei, K.~Goldstein and S.~Mahapatra,
  ``Moduli and (un)attractor black hole thermodynamics,''
  arXiv:hep-th/0611140.
  %%CITATION = HEP-TH/0611140;%%
  
  %\cite{Goldstein:2005rr}
\bibitem{Goldstein:2005rr}
  K.~Goldstein, R.~P.~Jena, G.~Mandal and S.~P.~Trivedi,
  ``A C-function for non-supersymmetric attractors,''
  JHEP {\bf 0602}, 053 (2006)
  [arXiv:hep-th/0512138].
  %%CITATION = JHEPA,0602,053;%%
  
 %\cite{Wiltshire:1985us}
\bibitem{Wiltshire:1985us}
  D.~L.~Wiltshire,
  ``Spherically Symmetric Solutions Of Einstein-Maxwell Theory With A
  Gauss-Bonnet Term,''
  Phys.\ Lett.\  B {\bf 169}, 36 (1986).
  %%CITATION = PHLTA,B169,36;%%

%\cite{Wiltshire:1988uq}
\bibitem{Wiltshire:1988uq}
  D.~L.~Wiltshire,
  ``Black Holes in String Generated Gravity Models,"
   Phys.\ Rev.\  D {\bf 38}, 2445 (1988).
  %%CITATION = PHRVA,D38,2445;%%
  
%\cite{Neupane:2002bf}
\bibitem{Neupane:2002bf}
  I.~P.~Neupane,
  ``Black hole entropy in string-generated gravity models,''
  Phys.\ Rev.\  D {\bf 67}, 061501 (2003)
  [arXiv:hep-th/0212092].
  %%CITATION = PHRVA,D67,061501;%%
  
    %\cite{Neupane:2003vz}
\bibitem{Neupane:2003vz}
  I.~P.~Neupane,
  ``Thermodynamic and gravitational instability on hyperbolic spaces,''
  Phys.\ Rev.\  D {\bf 69}, 084011 (2004)
  [arXiv:hep-th/0302132].
  %%CITATION = PHRVA,D69,084011;%%
  
 %\cite{Nampuri:2007gv}
\bibitem{Nampuri:2007gv}
  S.~Nampuri, P.~K.~Tripathy and S.~P.~Trivedi,
  ``On The Stability of Non-Supersymmetric Attractors in String Theory,''
  arXiv:0705.4554 [hep-th].
  %%CITATION = ARXIV:0705.4554;%%


\end{thebibliography}
\end{document}